# Competition between the Spin and Pseudospin Subsystems in a Model Cuprate

Yu. D. Panov[a,*], V. A. Ulitko[a], K. S. Budrin[a], D. N. Yasinskaya[a], and A. A. Chikov[a]

[a] *Ural Federal University Named after the First President of the Russian Federation B.N. Yeltsin, Yekaterinburg, 620002 Russia*

*\*e-mail: yuri.panov@urfu.ru*

**Abstract**—The competition between the magnetic and charge orderings in a model cuprate is considered in terms of a simplified static 2D spin–pseudospin model. This model is equivalent to the 2D dilute antiferromagnetic (AFM) Ising model with charged impurities. The mean-field approximation results are presented for the system under study and briefly compared to the classical Monte Carlo (MC) calculations. The numerical simulation shows that the cases of the strong exchange and the strong charge correlation differ qualitatively. In the case of a strong exchange, the AMF phase is instable with respect to the phase separation (PS) into the pseudospin (charge) and magnetic (spin) subsystems that behave as immiscible quantum liquids. The analytical expression has been obtained for the PS temperature.

## 1. INTRODUCTION

A topical problem of the physics of superconducting cuprates is the coexistence and the competition of the spin, superconducting, and charge orderings. The correlation between the magnetism and the superconductivity in cuprates has been studied for a long time [1, 2]. For the last fifteen years, many experimental results indicating the existence of the charge ordering [3–8] and the mutual influence of the spin and charge orderings in cuprates [9–14] were obtained. The model that relates the unique properties of cuprates to their instability to the charge transfer of the $CuO_4$ center states in the $CuO_2$ planes was proposed in [15]. This model makes it possible to consider, for $CuO_4$ centers in the $CuO_2$ plane, three many-electron valence states $CuO_4^{7-,6-,5-}$ (that formally correspond to the states of $Cu^{1+,2+,3+}$ cooper ions) as components of the pseudospin triplet $S = 1$ with $M_S = -1, 0, +1$, respectively, and enables us to use the pseudospin formalism for pseudospin $S = 1$ [15, 16]. To consider the competition between the spin and charge orderings in cuprates, the simplified static 2D spin–pseudospin model that is a limiting case of the general pseudospin model was proposed in [17–19].

Hamiltonian of a static spin–pseudospin model is

$$\mathcal{H} = \Delta \sum_i S_{zi}^2 + V \sum_{\langle ij \rangle} S_{zi} S_{zj} + \tilde{J} \sum_{\langle ij \rangle} \sigma_{zi} \sigma_{zj} \\ - \tilde{h} \sum_i \sigma_{zi} - \mu \sum_i S_{zi}, \quad (1)$$

where $S_{zi}$ is the $z$ component of pseudospin $S = 1$ on a site, and $\sigma_{zi} = P_{0i} s_{zi}/s$ is the normalized $z$ component of spin $s = 1/2$ multiplied by the projection operator $P_{0i} = 1 - S_{zi}^2$. Parameters $\Delta = U/2$ and $V > 0$ determine the on-site and the inter-site density-density interactions, respectively; $J = \tilde{J}/s^2 > 0$ is the Ising exchange interaction between $Cu^{2+}$ ions, $h = \tilde{h}/s$ is external magnetic field, $\mu$ is the chemical potential necessary for the inclusion of the condition of the doping charge constancy, and $nN = \sum \langle S_{zi} \rangle$ = const, where $n$ is the doping charge density. The summation is carried out over the 2D square lattice and $\langle ij \rangle$ implies the nearest neighbors. This spin–pseudospin model generalizes the 2D dilute antiferromagnetic (AFM) Ising model with charged impurities. In the limit $\Delta \to -\infty$, it reduces to the Ising model for spin $S = 1/2$ with a fixed magnetization. At $\Delta > 0$, the results can be compared to the Blume–Capel [20, 21] or to the Blume–Emery–Griffiths [22] model. The Ising model with mobile charged impurities was also considered in [23]. It is apparent that the most important restriction of our model for its comparison with real cuprates is the absence of charge transfer in the Hamiltonian.

The phase diagrams of the ground state were considered in the mean-field approximation in [17, 18]. It was shown that in all five phases of the ground state are realized in two limits. In the weak exchange limit at $\tilde{J} < V$, all the ground state phases (COI, COII, COIII, FIM) correspond to the charge ordering (CO) of a checker-board type at average charge density $n$. Whereas there are no spin centers ($Cu^{2+}$) in phase COI, phases COII and COIII are diluted with noninteracting spin centers distributed only in one sublattice. Such a ferrimagnetic spin ordering is a result of the mean-field approximation; because of this, the calculations by the classical Monte Carlo (MC) method in these cases show a paramagnetic response at low temperatures. The FIM phase is also formally ferrimagnetic. In this case, the spin AFM ordering is diluted with noninteracting charge centers ($Cu^{1+, 3+}$) distributed only in one sublattice. In the limit of strong exchange, at $\tilde{J} > V$, we observe only COI and AFM phases in which charge centers are homogeneously distributed in both sublattices.

This report is organized as follows. We present the results of the calculation of the thermodynamic properties of the system under study in the mean-field approximation and concisely compare them with the calculations by the classical MC method in Section 2. The MC calculations show that, in the limit of the strong exchange, the AFM phase is instable with respect to the phase separation (PS) into the subsystems of charge and spin centers. In Section 3, we analyze the thermodynamic properties of the PS state in a framework of coexistence of two homogeneous phases. Section 4 presents the conclusions.

## 2. MEAN-FIELD APPROXIMATION

In this section, we briefly present the results of the calculations of the thermodynamic properties in a mean-field (MF) approximation. We use the Bogolyubov inequality for the grand potential $\Omega(\mathcal{H})$: $\Omega(\mathcal{H}) \leq \Omega = \Omega(\mathcal{H}_0) + \langle \mathcal{H} - \mathcal{H}_0 \rangle$. In the standard way, we introduce two sublattices $A$ and $B$ on a square lattice and choose

$$\beta \mathcal{H}_0 = \delta \sum_i S_{zi}^2 - \sum_{\alpha, i_\alpha} \beta_\alpha S_{zi_\alpha} - \sum_{\alpha, i_\alpha} \gamma_\alpha \sigma_{zi_\alpha}, \quad (2)$$

where $\beta = 1/T$, $\delta = \beta\Delta$, $\beta_\alpha$ and $\gamma_\alpha$ are the molecular fields, $\alpha = A, B$. We obtain the expression for estimation of $\omega = \Omega/N$

$$2\beta\omega = \sum_\alpha [(\beta_\alpha - \xi)S_\alpha + (\gamma_\alpha - \eta)\sigma_\alpha \\ - \ln 2(e^{-\delta} \cosh\beta_\alpha + \cosh\gamma_\alpha)] + z\nu S_A S_B + zj\sigma_A\sigma_B, \quad (3)$$

where $\xi = \beta\mu$, $\nu = \beta V$, $j = \beta\tilde{J}$, $\eta = \beta\tilde{h}$, $z = 4$ is the number of the nearest neighbors, and the average (pseudo)magnetizations for sublattices $\langle S_{zi} \rangle_\alpha = S_\alpha$ and $\langle \sigma_z \rangle_\alpha = \sigma_\alpha$ have the form

$$S_\alpha = \frac{\sinh\beta_\alpha}{\cosh\beta_\alpha + e^\delta \cosh\gamma_\alpha}, \\ \sigma_\alpha = \frac{\sinh\gamma_\alpha}{e^{-\delta}\cosh\beta_\alpha + \cosh\gamma_\alpha}. \quad (4)$$

Minimizing $\omega$ with respect to $\beta_\alpha$ and $\gamma_\alpha$, we obtain the system of the MF equations

$$\beta_\alpha - \xi = -z\nu S_{\bar\alpha}, \quad \gamma_\alpha - \eta = -zj\sigma_{\bar\alpha}, \quad (5)$$

where $\bar{A} = B$ and $\bar{B} = A$.

Equations (5) must be complemented by charge restriction $S_A + S_B = 2n$. To explicitly include this condition, we can introduce the charge order parameter $a = (S_A - S_B)/2$ and to write the free energy $f = \omega + \mu n$ as a function of $n$, $a$, and $\sigma_\alpha$ using the inverse relationships for Eqs. (4):

$$e^{2\beta_\alpha} = \frac{(S_\alpha e^\delta + G_\alpha)^2 - \sigma_\alpha^2 e^{-2\delta}}{(1 - S_\alpha)^2 - \sigma_\alpha^2}, \\ e^{2\gamma_\alpha} = \frac{(\sigma_\alpha e^{-\delta} + G_\alpha)^2 - S_\alpha^2 e^{2\delta}}{(1 - \sigma_\alpha)^2 - S_\alpha^2}, \quad (6)$$

where

$$G_\alpha = (1 - S_\alpha^2 - \sigma_\alpha^2 + S_\alpha^2 e^{2\delta} + \sigma_\alpha^2 e^{-2\delta})^{1/2}.$$

For the nonordered (NO) high-temperature solution at $h = 0$, we have $a = 0$ and $\sigma_\alpha = 0$, and the free energy per one site takes the form

$$f_{\text{NO}} = \frac{z}{2}Vn^2 + \Delta|n| - \frac{1}{\beta}\ln\left(2\frac{1 + g_0}{1 - n^2}\right) \\ + \frac{|n|}{\beta}\ln\left(\frac{|n| + g_0}{1 - |n|}\right), \quad (7)$$

where $g_0 = ((1 - n^2)e^{-2\delta} + n^2)^{1/2}$. This enables us to calculate all the thermodynamic functions of the NO phase. The entropy, the internal energy, and the specific heat per one site are

$$s_{\text{NO}} = \delta\frac{(1 - |n|)(g_0 - |n|)}{1 + g_0} + \ln\left(2\frac{1 + g_0}{1 - n^2}\right) \\ - |n|\ln\left(\frac{|n| + g_0}{1 - |n|}\right), \quad (8)$$

$$e_{\text{NO}} = \frac{z}{2}Vn^2 + \Delta\frac{n^2 + g_0}{1 + g_0}, \quad (9)$$

$$c_{\text{NO}} = \delta^2\frac{(1 - n^2)^2 e^{-2\delta}}{g_0(1 + g_0)^2}. \quad (10)$$

Equations (7)–(10) correspond to the thermodynamic characteristics of the ideal system of noninteracting pseudospin (charge) and spin doublets separated in energy by the value $\Delta$ up to the temperature independent term $\frac{z}{2}Vn^2$. At $\Delta = 0$, the entropy and the internal energy become constant; thus, the specific heat is a zero. If $\Delta \neq 0$, the specific heat has a maximum at $T \propto |\Delta|$. In particular, if $n = 0$, then

$$c_{NO} = \left(\frac{\delta}{2}\right)^2 \cosh^{-2}\frac{\delta}{2}, \quad (11)$$

and the maximum is in point $T = |\Delta|/(2x)$, where $x$ is the root of equation $x = \coth x$.

We can also write the explicit form of the magnetic susceptibility at $h = 0$ in the NO phase. Assuming that $S_A = S_B = n$ and $\sigma_A = \sigma_B = \sigma$ at $h \neq 0$, we eliminate $\xi$ from system (5) and obtain equation

$$\sigma = \psi(\eta - xj\sigma, n), \quad (12)$$

where we introduced the following notations:

$$\psi(x, n) = \frac{(1 - n^2)\sinh x}{\cosh x + g(x, n)}, \quad (13)$$

$$g(x, n) = ((1 - n^2)e^{-2\delta} + n^2 \cosh^2 x)^{1/2}, \quad (14)$$
$$g(0, n) = g_0.$$

After the standard calculations, we obtain

$$\chi_{NO}|_{h=0} = \beta s^2 \left.\frac{\partial \sigma}{\partial \eta}\right|_{\eta=0} = s^2 \frac{\chi_0(n)}{1 + z\tilde{J}\chi_0(n)},$$
$$\chi_0(n) = \beta \frac{1 - n^2}{1 + g_0}, \quad (15)$$

where $\chi_0(n)$ is the normalized susceptibility in a zero external field for the ideal system of noninteracting pseudospin and spin doublets. System of equations (5) has the ferrimagnetic solutions with $\sigma_A + \sigma_B \neq 0$ at $h = 0$ [17] that are a result of the MF approximation and do not appear in MC calculations. Due to the short-range character of the exchange interaction in our model, these solutions can appear upon a numerical simulation as a mixture of the antiferromagnetic and paramagnetic phases. The underestimating of the paramagnetic response is the systematic error of the MF method in these cases. In what follows, we will consider only the AFM types of solutions with $\sigma_A = -\sigma_B = \sigma$ at $h = 0$. In this case, according to Eqs. (5), relationship $\gamma_A = -\gamma_B$ is fulfilled. The magnetizations of sublattices $\sigma_\alpha$ are monotonic functions of molecular fields $\gamma_\alpha$, according to Eqs. (4), therefore, only the case $\beta_A = \pm\beta_B$ is possible for $\sigma \neq 0$. This implies that, at $n \neq 0$, there are only AFM solutions with $a = 0$, $\sigma \neq 0$ and CO solutions with $a \neq 0$ and $\sigma = 0$. The case $n = 0$ should be considered individually, since it gives a possibility for frustrated states as the CO- and AFM-types solutions become degenerate.

The thermodynamic properties of the AFM and CO phases suggest knowledge of the roots of Eqs. (5) and can be calculated numerically. In addition, the equations for the second-order phase transition temperatures and the critical points can be found analytically.

For the AFM phase, we use condition $\partial^2 f/\partial\sigma^2 = 0$ at $\sigma = 0$, which gives

$$\left.\frac{\partial\gamma_\alpha}{\partial\sigma_\alpha}\right|_{\sigma_\alpha=0} = zj. \quad (16)$$

With accounting for Eqs. (6), we obtain the equation for the NO–AFM transition temperature

$$(1 - n^2)zj = 1 + g_0. \quad (17)$$

In particular, we obtain for $\Delta \to +\infty$ that

$$T_{AFM} = (1 - |n|)z\tilde{J}, \quad (18)$$

which coincides with the results of [22]. Substituting Eq. (17) into Eq. (15), we find the susceptibility in the transition point $\chi_{NO} = s^2/(2z\tilde{J})$.

To find the critical point that separates the first and second order transitions, we use equation $\partial^4 f/\partial\sigma_4 = 0$ in the coexistence curve. After some manipulations, we obtain

$$g_0^2 - 2g_0 - 3n^2 = 0. \quad (19)$$

Taking into account Eq. (17), we obtain the critical point position

$$T_{c1} = z\tilde{J}\frac{1 - n^2}{2 + \sqrt{1 + 3n^2}},$$
$$\frac{\Delta_{c1}}{T_{c1}} = \frac{1}{2}\ln\frac{1 - n^2}{2(1 + n^2 + \sqrt{1 + 3n^2})}. \quad (20)$$

In particular, at $n = 0$, $T_{c1} = z\tilde{J}/3$, $\Delta_{c1}/T_{c1} = -\ln 2$, which agrees with the results of [22].

The susceptibility in a zero field in the AFM phase has the form

$$\chi_{AFM}|_{h=0} = s^2 \frac{\beta\psi'(zj\sigma, n)}{1 + zj\psi'(zj\sigma, n)}, \quad (21)$$

where

$$\psi'(x, n) = \frac{(1 - n^2)(g(x, n) + g_0^2 \cosh x)}{g(x, n)(\cosh x + g(x, n))^2}. \quad (22)$$

The order parameter of the AFM phase $\sigma$ at $h = 0$ can be found from equation

$$\sigma = \psi(zj\sigma, n). \quad (23)$$

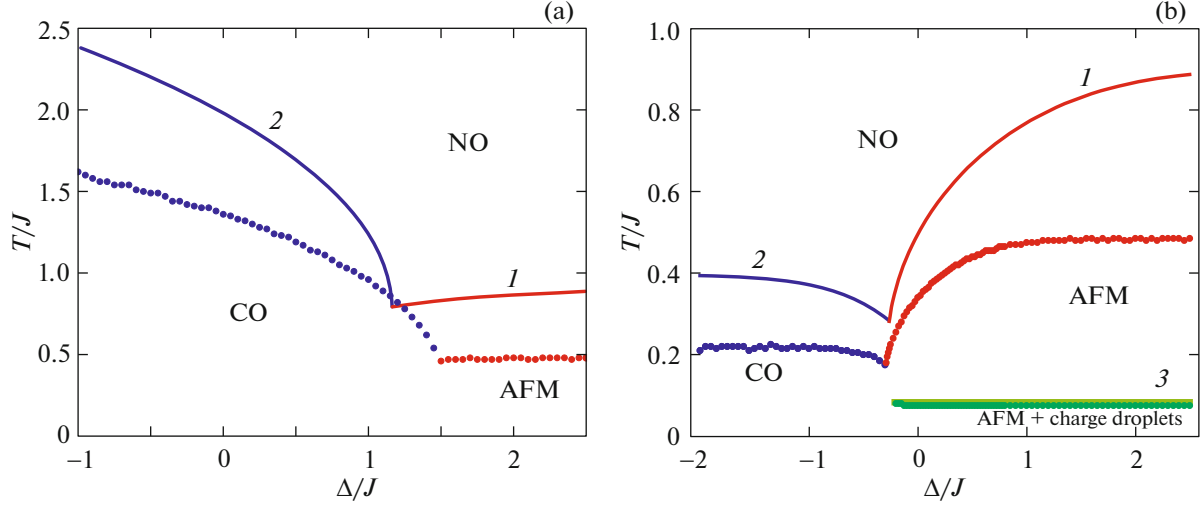

**Fig. 1.** (a) Left panel: the case of weak exchange: $n = 0.1$, $\tilde{J} = 0.25$, and $V = 1$; (b) the case of the strong exchange: $n = 0.1$, $\tilde{J} = 0.25$, and $V = 0.1$. Points correspond to the MC critical temperatures. The values of the MF critical temperature given by Eqs. (17), (25), and (34) are indicated in lines *1–3*.

By analogy, for the CO phase, condition $\partial^2 f/\partial a^2 = 0$ at $a = 0$ gives

$$\frac{1}{2}\left(\frac{\partial \gamma_A}{\partial a} - \frac{\partial \gamma_B}{\partial a}\right)\bigg|_{\sigma_\alpha=0} = z\nu, \quad (24)$$

and we obtain equation for the NO–CO transition temperature

$$(1 - n^2)z\nu = 1 + g_0^{-1}. \quad (25)$$

In particular, for $\Delta \to -\infty$, we obtain

$$T_{CO} = (1 - n^2)zV. \quad (26)$$

In the CO phase, the equation for the critical point is more complex

$$2(1 + 3n^2)g_0^3 - g_0^2 - 6n^2 g_0 + 3n^2 = 0, \quad (27)$$

but, at $n = 0$, this gives $T_{c2} = zV/3$, $\Delta_{c2}/T_{c2} = \ln 2$.

In a zero field the susceptibility in the CO phase is

$$\chi_{CO}\big|_{h=0} = s^2 \frac{\frac{1}{2}(\chi_0(n+a) + \chi_0(n-a))}{1 + \frac{1}{2}z\tilde{J}(\chi_0(m+a) + \chi_0(n-a))}, \quad (28)$$

where the CO-phase order parameter obeys equation

$$a = \frac{1}{2z\nu}\ln\left(\frac{(n+a+g(0,n+a))(1-n+a)}{(n-a+g(0,n-a))(1-n-a)}\right). \quad (29)$$

The general formula for the susceptibility in a zero field combining the cases of the NO, AFM, and CO phases is given by relationship

$$\chi = s^2 \frac{\frac{1}{2}\beta(\psi'(zj\sigma, n+a) + \psi'(zj\sigma, n-a))}{1 + \frac{1}{2}zj(\psi'(zj\sigma, n+a)(+\psi'(zj\sigma, n-a))}, \quad (30)$$

where $\sigma$ is the AFM-phase order parameter, and $a$ is the CO-phase order parameter.

The numerical simulation was carried out using a high-efficient algorithm of parallel calculations and the classical MC method. Figure 1 shows the results of the MC calculations. The peak position in the temperature dependence of the specific heat approximately (because of a finite size of the system) corresponds to the disorder–order transition temperature. These values are indicated by points. The transition temperatures obtained in the MF approximation are shown by solid lines. Figure 1 clearly demonstrates the typical (by a factor of slightly lower than two times) overestimation of the critical temperature value in the MF approximation. Figure 3 shows the dependence of the AFM-transition critical temperature on $n$ in the MF approximation. Its shape qualitatively agrees with the results obtained in the Bethe approximation in [23].

Figure 2 presents the comparison of the results for the susceptibility and the specific heat obtained by the MF approximation and by MC method. The analytical MF-dependences show a qualitative agreement to the results of the numerical simulation and, in some cases, even a quantitative coincidence of the results for the high-temperature region. The main discrepancies are due to the difference in the critical temperature and systematic inaccuracies in the MF approximation in the case of description of the critical fluctuations and the paramagnetic response at low temperatures.

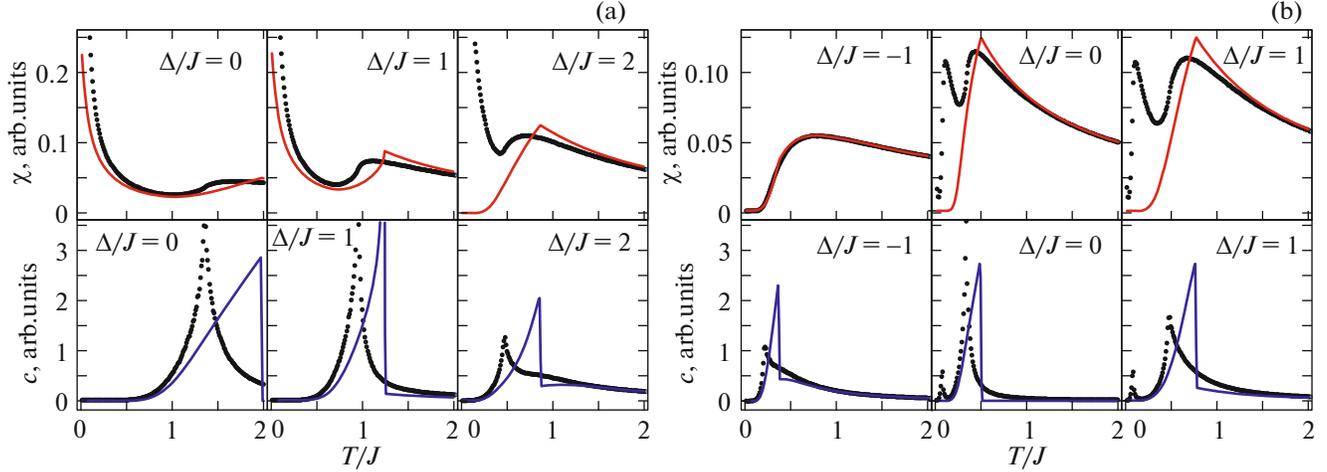

**Fig. 2.** Susceptibility and the specific heat obtained (solid lines) in the MF approximation and (points) by MC method: (a) the case of weak exchange: $n = 0.1$, $\tilde{J} = 0.25$, and $V = 1$ and (b) the case of the strong exchange: $n = 0.1$, $\tilde{J} = 0.25$, and $V = 0.1$.

## 3. THE PHASE SEPARATION CRITICAL TEMPERATURE

According to the results of the numerical simulation by the MC method, the temperature dependences of the specific heat demonstrate in the limit of strong exchange at positive $\Delta$ two successive phase transitions. The immediate study of the system state shows that the first transition is the AFM ordering. With lowering temperature in the spin subsystem that is an AFM matrix diluted by randomly distributed charged impurities, the impurities are condensed into charge droplets. It means that in the limit of the strong exchange, the diluted AFM-phase is instable with respect to the macroscopic (PS) into the pseudospin (charge) and magnetic (spin) subsystems. At this stage, the AFM matrix forces out charged nonmagnetic impurities to minimize the surface energy. Note that, in the limit of weak exchange, the charged impurities are randomly distributed over the AFM matrix up to $T = 0$ and also the charged impurities are randomly distributed in the CO phase, since the energies of all possible distributions of additional charges over the CO matrix are the same in the approximation of interaction of only the nearest neighbors. The results of our numerical simulation are similar to the results obtained for binary alloys in [24, 25].

To describe the thermodynamic properties of the heterogeneous state, we use the model developed in [26–28] for a macroscopic PS in electron systems. This model is based on the Maxwell construction. Assuming that there are two coexisting macroscopic homogeneous phases 1 and 2, we write the free energy of the PS state per one site as

$$f_{PS} = m f_1(n_1) + (1-m) f_2(n_2), \quad (31)$$

where $m$ is the system fraction with density $n_1$, $1 - m$ is the system fraction with density $n_2$, so that $m n_1 + (1-m) n_2 = n$. In our case, one phase consists of charged centers (C), and another phase is the spin AFM phase without impurities; therefore, $n_1 = \text{sgn} n$, $n_2 = 0$, and $m = |n|$. The transition point is determined by equation

$$|n| f_C(1) + (1-|n|) f_{AFM}(0) = f_{AFM}(n). \quad (32)$$

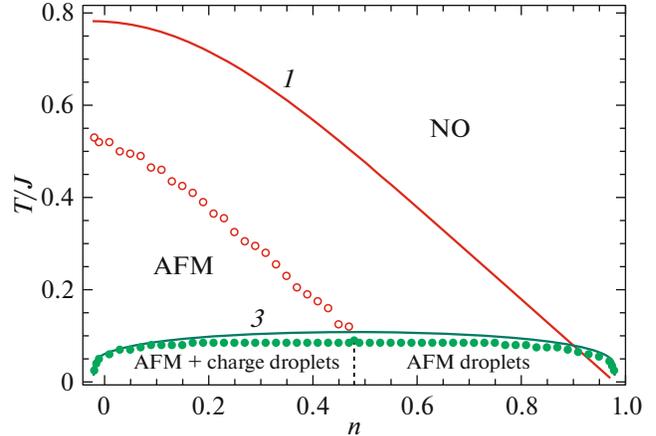

**Fig. 3.** The bright circles correspond to the susceptibility maxima upon the AFM ordering; dark cycles correspond to the specific heat maxima upon PS obtained by the MC method. The model parameters are $\Delta = 1$, $\tilde{J} = 0.25$, and $V = 0.14$. Lines *1* and *3* show the critical temperatures given by Eqs. (17) and (34).

The free energy of charge centers $f_C(1) = 2V + \Delta$. The free energy of the AFM phase can be written as

$$f_{\text{AFM}}(n) = \frac{z}{2}(Vn^2 + \tilde{J}\sigma^2) + |n|\Delta \\ - \frac{1}{\beta}\ln\left(2\frac{\cosh(zj\sigma) + g(zj\sigma, n)}{1 - n^2}\right) \\ + \frac{|n|}{\beta}\ln\left(\frac{|n|\cosh(zj\sigma) + g(zj\sigma, n)}{1 - |n|}\right). \quad (33)$$

We assume that $\Delta > 0$ and consider the case of low temperatures, thus, $\delta \gg 1$ and $j \gg 1$. In this approximation, with the inclusion of Eq. (23), we obtain $|\sigma| = 1 - |n|$. As a result, Eq. (32) gives the following relationship for PS temperature

$$T_{\text{PS}} = \frac{|n|(1-|n|)}{|n|\ln|n| + (1-|n|)\ln(1-|n|)}\frac{z(V - \tilde{J})}{2}. \quad (34)$$

This expression does not depend on $\Delta$, which agrees with the MC results for $T_{\text{PS}}$ in Fig. 1.

Figure 3 shows the concentration dependences of the critical temperatures for the AFM transition and the PS. The circles denote the MC results for the susceptibility maxima upon the AFM ordering, and the points show the specific heat maxima upon the PS. Solid line *3* shows the PS temperature determined by Eq. (34). This temperature agrees unexpectedly well with the MC results, while the dependence in the MF approximation (line *2*) for the AFM-ordering temperature determined by Eq. (17) becomes qualitatively wrong at $|n| > 0.5$. Note as well that Eq. (34) for the PS temperature based on the Maxwell construction gives lower values as compared to the values found in [22].

## 4. CONCLUSIONS

We considered the static 2D spin–pseudospin model on a square lattice that generalizes the dilute antiferromagnetic Ising model. We compared similar results in the MF approximation with the results of the numerical simulation by the classical MC method. The analysis of the specific heat and the susceptibility obtained by the MC method showed that the MF critical temperature of CO and AFM ordering qualitatively reproduce the numerical results, but they systematically give higher values. The MC calculations show that the cases of the strong and weak exchange are qualitatively different. In the case of the weak exchange, a frustration appears in the charge-ordered ground state of the system. In the limit of the strong exchange, the homogeneous AFM phase is instable with respect to the PS of the pseudospin and spin subsystems. We obtained the analytical expression for the PS temperature and revealed that it agrees well with the numerical simulation by the classical MC method.

## FUNDING

This work was supported by Program 211 of the Government of the Russian Federation (Agreement 02.A03.21.0006), the Ministry of Education and Science of the Russian federation (projects nos. 2277 and 5719), and the Russian Foundation for Basic Research (project no. 18-32-00837\18).